\documentclass[twocolumn,preprintnumbers,amsmath,aps,longbibliography]{revtex4-2}
\usepackage{graphicx}
\usepackage{dcolumn}
\usepackage{bm}
\usepackage{subfigure}
\usepackage{color}
\usepackage{amssymb}
\begin{document}
\newcommand{\kvec}{\mbox{{\scriptsize {\bf k}}}}
\def\eq#1{(\ref{#1})}
\def\fig#1{figure\hspace{1mm}\ref{#1}}
\def\tab#1{table\hspace{1mm}\ref{#1}}
\title{Interplay of Kekul{\' e} order, flat bands and electron-phonon coupling in graphene}
\author{Dominik Szcz{\c{e}}{\'s}niak}\email{d.szczesniak@ujd.edu.pl}
\affiliation{Institute of Physics, Faculty of Science and Technology, Jan D{\l}ugosz University in Cz{\c{e}}stochowa, 13/15 Armii Krajowej Ave., 42200 Cz{\c{e}}stochowa, Poland}
\date{\today} 
\begin{abstract}

The Kekul{\'e} order in graphene is known to coexist with both an extended flat band near the van Hove singularity (VHS) and enhanced electron-phonon coupling. However, the microscopic origin of this interplay is still not fully understood. Here, this behavior is traced to the Kekul{\'e}-induced bond distortion and the resulting redistribution of electronic delocalization. In this picture, the conduction band below the VHS is observed to flatten with increasing Kekul{\'e} order, extending its intrinsically weakly dispersive region around the saddle point over a broader momentum range. In parallel, the VHS and dimensionless electron-phonon coupling exhibit approximately quadratic upward responses. The resulting enhancement of the electron-phonon coupling is found to be predominantly governed by the increasing density of states, with the effective deformation potential remaining weakly affected by the distortion. As such, the obtained results are consistent with previous theoretical and experimental studies, while also pointing toward the narrowing of the conduction band and a possible departure of the electron-phonon coupling from the conventional adiabatic regime. Altogether, they highlight the potential of controlled Kekul{\'e} ordering for engineering emergent correlated quantum phases in graphene.

\end{abstract}
\maketitle
%

\section{Introduction}

The Kekul{\'e} order can be associated with a variety of intriguing phenomena that are absent in pristine graphene. These include, but are not limited to, the band gap opening at the Dirac point \cite{hou2007, lin2017, bao2021}, valley mixing \cite{ryu2009, gutierrez2016}, as well as the emergence of valence bond ordering \cite{naumis2024, costa2024} and superconductivity \cite{ludbrook2015, szczesniak2019}. Notably, Kekul{\'e} order is also known to coexist with a flat band that forms near the van Hove singularity (VHS) above the Dirac point \cite{bao2022, zaarour2023, qiu2026}. This band not only suppresses the electronic kinetic energy but also enhances many-body interactions in graphene, which can dominate its low-energy physics and stabilize new correlated quantum phases. Of particular interest in this context are electron-phonon interactions, as they drive Kekul{\'e} ordering \cite{gutierrez2016}, become further amplified by the resulting flat band \cite{bao2022}, and promote notable instabilities like conventional superconductivity \cite{profeta2012}.

Although the above relations are known, a unified description linking Kekul{\'e} order, flat band formation, and the resulting enhancement of electron-phonon coupling in graphene is still lacking. This concerns not only the reason why the Kekul{\'e} instability coexists with flat bands but also how it gives rise to the observed strong band renormalization effects. So far, the discussed interplay has been supported mostly by empirical observations \cite{bao2022, zaarour2023}. In contrast, the present analysis aims to address it directly at the microscopic level. For this purpose, it is recalled that structural perturbations intrinsically alter the atomic environment in which electrons respond to lattice displacements, thereby also modifying the electron-phonon interaction \cite{gustino2017}. In this picture, the related strength of the electron-phonon coupling is given by the fundamental Hopfield-McMillan formula \cite{mcmillan1968, hopfield1969}:
\begin{equation}
\label{eq01}
\lambda = \frac{N(E_{F}) D^2}{M \Omega^2},
\end{equation}
where $N(E_{F})$ denotes the density of states at the Fermi energy ($E_F$), $D$ stands for the deformation potential, while $M$ and $\Omega$ describe the atomic mass and phonon frequency, respectively. Interestingly, in graphene only the two former quantities are expected to change notably due to the Kekul{\'e} ordering \cite{bao2022, costa2024}. Moreover, for the dominant in-plane phonon mode ($A_{1}$), their microscopic behavior can be traced down to the level of individual bonds and the electronic response to lattice displacements, as expressed by \cite{sheng2015, stauffert2019, silva2019, szczesniak2026}:
\begin{equation}
\label{eq02}
t \approx t_{0}+\alpha(u_{0}-u),
\end{equation}
with $t$ ($t_0$) describing distorted (equilibrium) hopping for $u$ ($u_{0}$) bond length, and $\alpha$ denoting corresponding bond-resolved electron-phonon coupling.

In the context of the above, both the density of states and the deformation potential appear to carry complementary information on the electronic and coupling properties of the Kekul{\'e} ordered system. This implies that they may also enable the promotion of the local bond-level picture to the microscopic framework that quantitatively links the effects of interest. Building on this perspective, the following analysis aims to establish a unified description that clarifies coexistence of lattice distortion, band renormalization, and enhanced electron-phonon interaction in Kekul{\'e} ordered graphene. This approach intends to provide not only a foundation for understanding the underlying mechanisms of this interplay, but also insight into their role in the emergence of corresponding correlated quantum phases.

\section{Kekul{\' e} order}

\begin{figure}[ht!]
\includegraphics[width=\columnwidth]{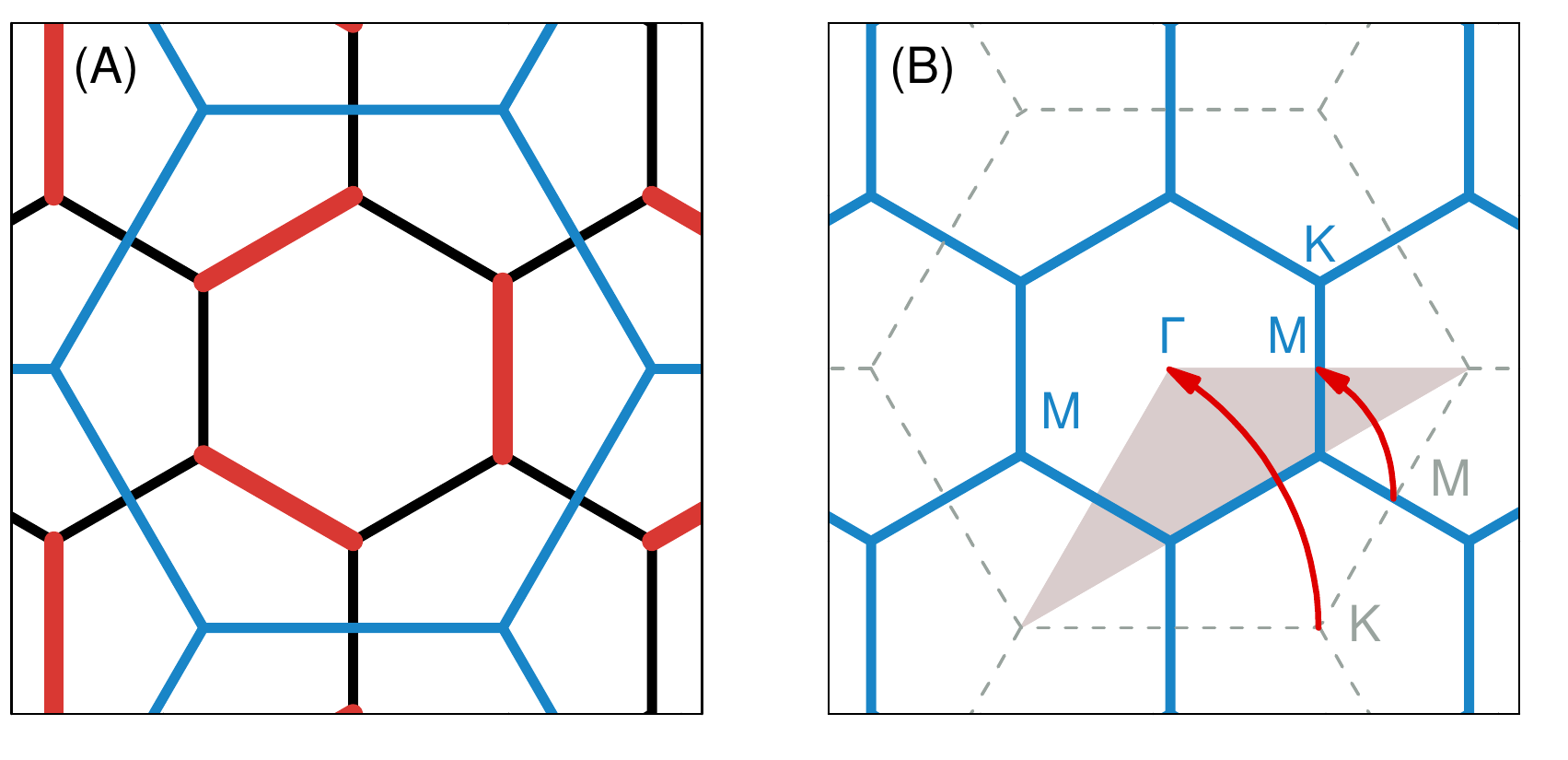}
\caption{(A) The $\left(\sqrt{3} \times \sqrt{3} \right) {\rm R} 30^{o}$ supercell of graphene (blue solid lines) along with schematics of the Kekul{\' e} distortion that introduces alternating shorter (red solid lines) and longer (black solid lines) bonds between carbon atoms. (B) The Brillouin zone folding of the original zone of graphene (dashed grey lines) onto its reduced counterpart (blue solid lines) together with the most characteristic high symmetry points.}
\label{fig01}
\end{figure}

The Kekul{\' e} order in graphene is characterized by an alternating bond length modulation, resulting in a $\left(\sqrt{3} \times \sqrt{3} \right) {\rm R} 30^{o}$ superlattice, as schematically illustrated in Fig. \ref{fig01} (A). Accordingly, the corresponding Brillouin zone, shown in Fig. \ref{fig01} (B), becomes reduced with respect to that of pristine graphene. In this representation, special attention is given to the new $\Gamma$ point, where the inequivalent $K$ and $K'$ valleys are folded, and to the midpoints of the reduced zone that coincide with the original $M$ points. These are the distinctive high-symmetry locations in the momentum space that host signatures of the Kekul{\'e} order relevant to the present study. In particular, the energy gap opens at the zone center, whereas the saddle-point states responsible for the VHS singularity are centered around the $M$ points \cite{cheianov2009, shao2015, qu2022, qiu2026}.

In this framework, the local insight encoded in Eq. (\ref{eq02}) is conveniently captured by a spinless Hamiltonian that includes hopping of $p_z$ electrons up to third-nearest neighbors. Such effective model reproduces all key features central to the present discussion, including the electron-hole asymmetry and the correct position and curvature of the VHS, taking the form \cite{szczesniak2026}:
\begin{equation}
\label{eq03}
H = \sum_i \epsilon c_i^\dagger c_i - \sum_{n=0}^{2} t_n \sum_{\langle i,j\rangle_n}
\left( c_i^\dagger c_j + c_j^\dagger c_i \right),
\end{equation}
where $\epsilon$ is the energy of a $p_{z}$ electron on site $i$ and $t_{n}$ denotes $n$-th hopping between $i$-th and $j$-th sites, with corresponding creation/annihilation operators $c_i$ and $c_j$. Here, the latter parameters are of particular interest as they are directly affected by the Kekul{\' e} bond modulations, governing band renormalization effects as well as the redistribution of spectral weight near the VHS. As a result, these parameters allows to introduce the Kekul{\' e} distortion into the Eq. (\ref{eq03}) through the following fundamental relation \cite{farjam2009, szczesniak2026}:
\begin{equation}
\label{eq04}
t_{0} = \frac{2t_{0}'  + t_{0}''}{3},
\end{equation}
with $t_{0}'$ and $t_{0}''$ being the nearest neighbor hoppings for the shorter two-thirds and longer one-third bonds, respectively. Note that the distortion described by Eq. (\ref{eq04}) alternates also the higher order hoppings through modification of corresponding distances between carbon atoms.

Based on the above, proxy for the magnitude of the Kekul{\' e} distortion can be defined by relating the unbalanced nearest neighbor hoppings to the energy gap induced as:
\begin{equation}
\label{eq05}
\Delta = 2|t_{0}''-t_{0}'|.
\end{equation}
This assumption is reinforced by the fact that in the Dirac description the Kekul{\' e} order enters as a mass term. This is to say, the gap size naturally scales with the amplitude of the distortion and vanishes when it disappears.

\section{Flat band}

In the spirit of Eq. (\ref{eq01}), it is instructive to first analyze interplay between Kekul{\' e} order and electronic properties of graphene through the density of states. This way, the pivotal coexistence of the ordered state and flat band can be addressed before establishing its implications for the electron-phonon coupling according to Eq. (\ref{eq02}). In particular, for the $\left(\sqrt{3} \times \sqrt{3} \right) {\rm R} 30^{o}$ superlattice the density of states at given energy ($E$) can be expressed as:
\begin{equation}
\label{eq06}
N(E)=\frac{1}{N}\sum_{\mathbf{k},n}\frac{1}{\sqrt{2\pi}\,\sigma_E}\exp\left[-\frac{\big(E-E_n(\mathbf{k})\big)^2}{2\sigma_E^{\,2}}\right],
\end{equation}
where $N=300^2$ is the total number of sampled momentum points in the employed $300\times300$ Brillouin-zone mesh, and $\sigma_E=30$~meV is the Gaussian broadening parameter. Moreover, $E_{n}(\mathbf{k})$ denotes the energy of the $n$-th electronic band at momentum $\mathbf{k}$, obtained from the eigenvalue problem corresponding to Eq. (\ref{eq03}) and accounting for orbital overlaps. Note that all the parameters entering the eigenvalue problem are adopted from \cite{gruneis2008} with the spatial dependence as introduced in \cite{szczesniak2026} for the graphene with Kekul{\' e} order.

\begin{figure*}[ht!]
\includegraphics[width=\textwidth]{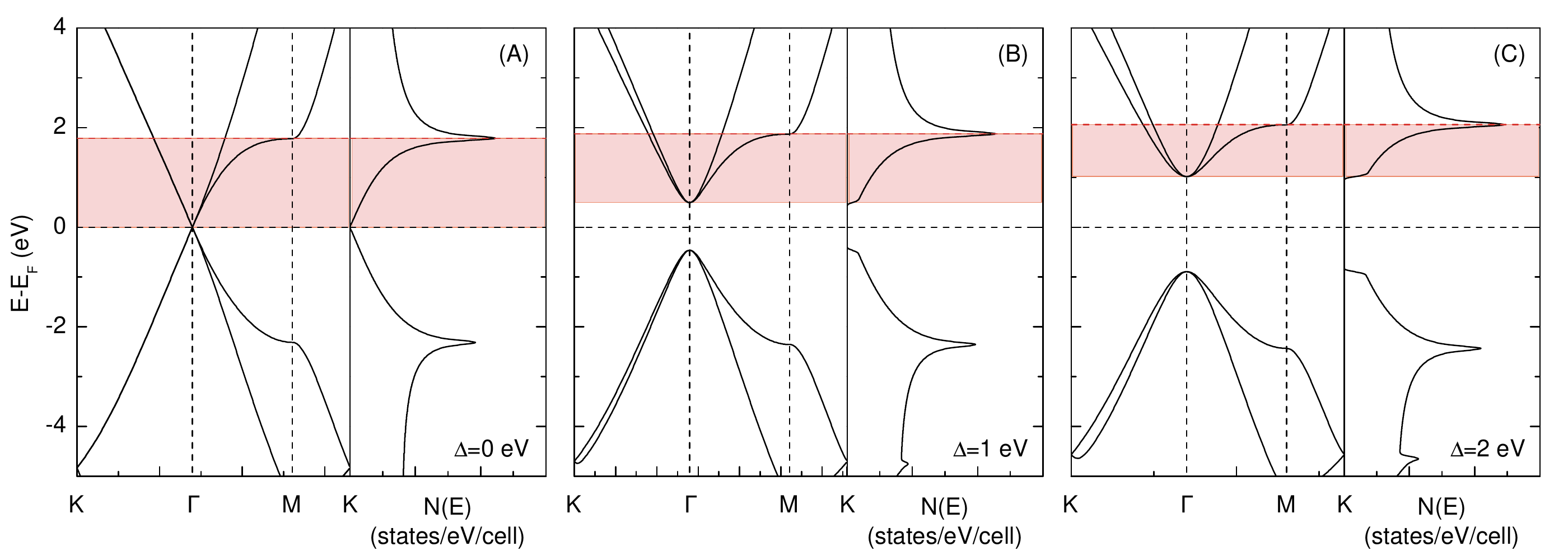}
\caption{The electronic structure and the related density of states of graphene for selected values of the band gap proxy. The zero reference level is set at the Fermi energy (black dotted line), so the maximum of the VHS associated with the conduction band lies at $E_{\rm VHS}\approx2$~eV (red dotted line). The conduction band width below $E_{\rm VHS}$ is marked by the red shaded area.}
\label{fig02}
\end{figure*}

\begin{figure}[ht!]
\includegraphics[width=\columnwidth]{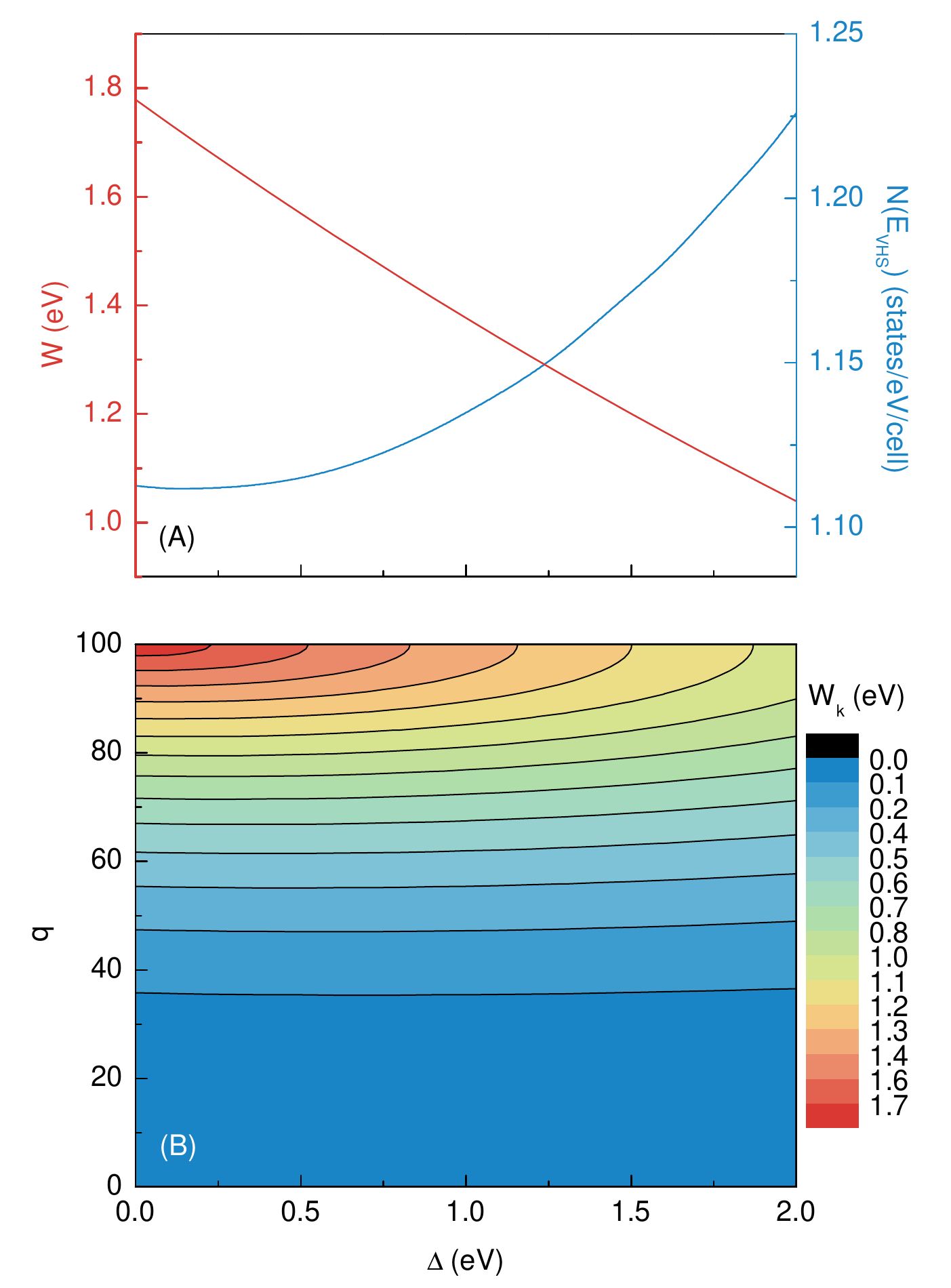}
\caption{(A) The conduction band width below $E_{\rm VHS}$ (red) and the maximum of the associated VHS (blue) for graphene as a function of the band gap proxy. (B) The momentum-resolved conduction band width as a function of the band gap proxy and the relative position along the $M-\Gamma$ path.}
\label{fig03}
\end{figure}

In Fig. \ref{fig02} the electronic structure of graphene and the density of states for selected values of the band gap proxy are presented. The band gap ranges from 0~eV to 2~eV in order to account for both the pristine graphene and representative distorted structures, including lithium-decorated graphene \cite{bao2021}, as well as more extreme cases considered before \cite{khoa2019, park2015}. For the undistorted case, the reference electronic structure, consisting of valence and conduction bands with the Dirac point at $E_{F}=0$~eV, is depicted in Fig. \ref{fig02} (A) and gives rise to two distinctive and asymmetric VHSs at approximately $\pm 2$~eV in the corresponding $N(E)$ spectrum. However, as the gap opens and increases in size, the electronic dispersion becomes visibly renormalized in Figs. \ref{fig02} (B) and (C). In particular, with the increasing magnitude of the Kekul{\' e} order, the bands become progressively compressed in energy and the three-fold degeneracy at the $K$ point is lifted, producing a small kink in the density of states. Yet, the most important effect is the simultaneous flattening of the conduction band around the $M$ point that leads to an enhancement of the associated VHS at $E_{\rm VHS}\approx2$~eV. This trend is best illustrated quantitatively by the joint evolution of the conduction band width ($W$) below $E_{\rm VHS}$ and the associated peak in the density of states spectrum ($N(E_{\rm VHS})$), as shown in Fig. \ref{fig03} (A). For this cross section of the conduction band, its width is reduced by about $41\%$, whereas the maximum of the VHS increases by $10\%$ in the maximally gapped structure compared with pristine graphene. Interestingly, while $W$ decreases almost linearly with the band gap, $N(E_{\rm VHS})$ exhibits an approximately quadratic dependence, indicating an increasingly pronounced spectral response to the progressive band flattening. Notably, this behavior departs from the trivial heuristic scaling $N(E_{\rm VHS})\sim 1/W$, indicating that the enhancement of the VHS cannot be attributed solely to the overall compression of the band. Instead, the different evolution of these observables points to a distortion-induced reshaping of the local conduction band dispersion.

In this context, Fig. \ref{fig03} (B) depicts the momentum-resolved conduction band width, given as:
\begin{equation}
\label{eq07}
W_{k}=E_{\rm VHS}-E_{\rm C}(\mathbf{k}),
\end{equation}
where $E_{\rm C}(\mathbf{k})$ is the conduction band energy at momentum $\mathbf{k}$. The evolution of $W_{k}$ is presented as a function of the band gap proxy and the normalized momentum coordinate:
\begin{equation}
\label{eq08}
q=\frac{|\mathbf{k}-\mathbf{k}_{M}|}{|\mathbf{k}_{\Gamma}-\mathbf{k}_{M}|},
\end{equation}
with $\mathbf{k}_{M}$ and $\mathbf{k}_{\Gamma}$ corresponding to the momentum coordinates of the $M$ and $\Gamma$ points, respectively. In this notation, $q=1$ denotes the Brillouin zone center, whereas $q=0$ refers to the saddle point location. The presented results clearly show that the influence of the Kekul{\' e} order on $W_k$ becomes stronger with increasing $q$ {\it i.e.} as $\mathbf{k}$ departs from the $M$ point. In detail, $W_k$ decreases by about $2\%$ even within the already relatively flat region at $q=0.2$, with this reduction progressively increasing to about $6\%$, $11\%$, $20\%$, and the aforementioned $41\%$ at $q=0.4$, $0.6$, $0.8$, and $1$, respectively. In other words, the Kekul{\'e} order not only flattens the conduction band but also extends its weakly dispersive region over a broader momentum range.

As such, the obtained results qualitatively reproduce, in the folded Brillouin-zone representation, the extended flat band observed experimentally for lithium-decorated graphene in \cite{bao2022}. In agreement with this study, the flat band originates from the saddle point region associated with the VHS, which can be experimentally approached through atom doping \cite{bao2021, bao2022}. For pristine graphene, this regime corresponds to the $1/4$ filling relative to charge neutrality within the nearest-neighbor approximation \cite{wang2012, gholami2018}, and is referred to here more generally as the van Hove filling. In this context, the presented observations show that the flat band and the Kekul{\' e} order may indeed codevelop in distorted graphene at the van Hove filling. However, the development of the extended flat band is in fact largely associated with the pronounced reduction of the conduction band width beyond the intrinsically weakly dispersive region in the vicinity of the saddle point. This suggests that the flattening is not fully explained by the processes responsible for establishing the ordered state. In particular, while the screened Coulomb repulsion between dopants and their interaction with the substrate drive the formation of the ordered state \cite{bao2022, song2012}, further flattening originates from the resulting bond-length modulation, which changes the overlap between neighboring $p_{z}$ orbitals and thus their degree of electronic delocalization. Consequently, this microscopic mechanism is captured here by the hopping parameters according to Eq. (\ref{eq04}), which effectively reduce width of the conduction band width below $E_{\rm VHS}$.

\section{Electron-phonon coupling}

To analyze how the joint effects of the Kekul{\' e} ordering and extended flat-band formation influence the electron-phonon coupling, an effective Hopfiled-McMillan formula is employed. In the framework adopted here, the corresponding dimensionless constant can be expressed for the dominant in-plane phonon mode ($A_{1}$) as:
\begin{equation}
\label{eq09}
\lambda \approx \frac{N(E_{\rm VHS})\alpha^2}{6 M_{C} \Omega_{A_{1}}^2},
\end{equation}
where the density of states is evaluated at the VHS ($N(E_{\rm VHS})$), to simulate van Hove filling and account for the most pronounced effect of the extended flat band on the $N(E)$ spectrum. Moreover, the deformation potential is argued to scale with the bond-resolved electron-phonon coupling ($D \propto \alpha$), since the lattice displacement in distorted graphene can be assumed to perturb the electronic states mainly through the modulation of hopping \cite{castro2007, cappelluti2012, szczesniak2026}. This allows to approximate deformation potential by $\alpha$, which effectively accounts for the contributions from the shorter two-thirds ($\alpha'$) and longer one-third ($\alpha''$) bonds through their bond-weighted root mean square:
\begin{equation}
\label{eq10}
\alpha=\sqrt{\frac{2\alpha'^2  + \alpha''^2}{3}},
\end{equation}
assuming that $\alpha'$ and $\alpha''$ are obtained from $\left|\partial t(u)/\partial u \right|$ evaluated at the corresponding displacements with $t(u)=t_{0}{\rm exp}[\eta(u/u_{0}-1)]$ and the phonon softening/hardening rate $\eta=2.13$ \cite{szczesniak2026}. The above reflects the fact that the electron-phonon matrix elements enter the Hopfield-McMillan expression quadratically, while the adopted averaging preserves the individual contributions of the two inequivalent bond classes. Finally, in Eq. (\ref{eq09}), $M_{C}=12m_{u}$ is the carbon atomic mass with $m_{u}=1.66054 \times 10^{-27}$ kg, while $\Omega_{A_{1}}$ denotes the characteristic frequency of the $A_{1}$ phonon mode, taken such that  $\hbar\Omega_{A_{1}}=170$ meV \cite{costa2024}.

\begin{figure}[ht!]
\includegraphics[width=\columnwidth]{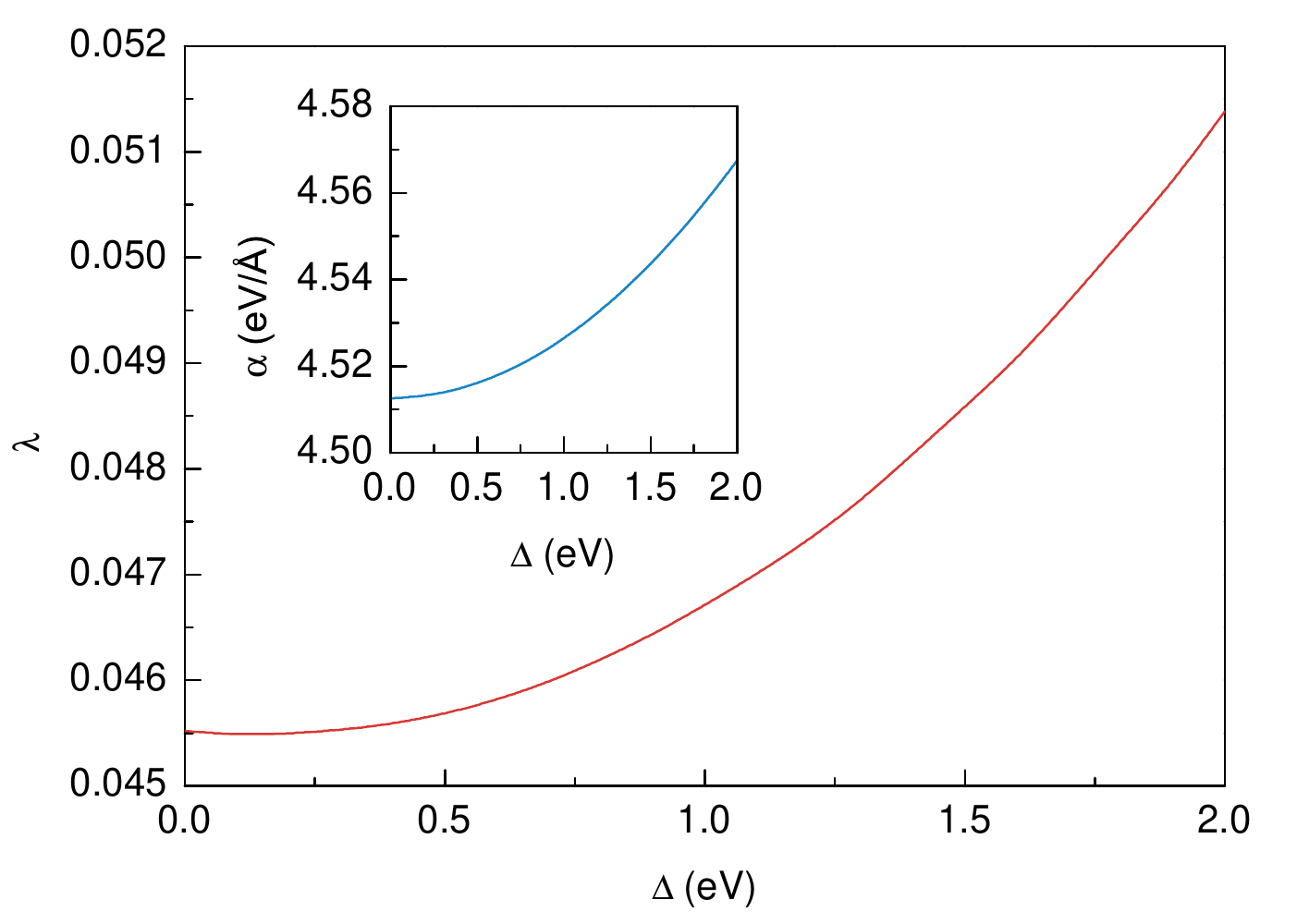}
\caption{The effective dimensionless electron-phonon coupling for the dominant in-plane phonon mode ($A_{1}$) of graphene as a function of the band gap proxy. The same functional dependence for the corresponding bond-resolved electron-phonon coupling is presented in the inset.}
\label{fig04}
\end{figure}

In Fig. \ref{fig04}, the dependence of the electron-phonon coupling on the band gap proxy is depicted, showing a notable strengthening of the former along with the increasing Kekul{\' e} order. Specifically, $\lambda$ is found to be almost $13\%$ higher for the graphene with largest considered distortion compared to pristine structure. Importantly, according to earlier analysis, the observed behavior of the electron-phonon coupling strongly correlates with the flattening of the conduction band. In fact, $N(E_{\rm VHS})$ which conveniently quantifies the flattening effect appears to have greater impact on the electron-phonon coupling than the effective $\alpha$. This is to say, when $N(E_{\rm VHS})$ increases by around $10 \%$ in Fig. \ref{fig03} (A), $\alpha$ experiences growth of only sightly more than 1 \%, as shown in the inset of Fig. \ref{fig04}. However, this effect becomes even more pronounced when recalling that the density of states is evaluated here at the VHS for all considered Kekul{\' e} order magnitudes. In particular, shifting the Fermi energy from the Dirac point toward the VHS already changes $N(E)$ from zero to the finite value, whereas no comparable modification is expected for $\alpha$. Hence, the electronic density of states appears to play the dominant role in shaping the electron-phonon coupling, both via filling-induced shift and its further enhancement with increasing Kekul{\' e} order, while the effective deformation potential remains almost constant. In general, these observations are in line with the signatures of strong electron-phonon coupling previously observed experimentally in the lithium-decorated graphene when the extended flat band is created due to the Kekul{\' e} order \cite{bao2022}. They also reinforce earlier qualitative reasoning that suggests potential of exploiting anisotropy of the bond-resolved electron-phonon coupling to enhance $\lambda$ in graphene through structural engineering \cite{szczesniak2026}. 

In addition to the above, it is informative to also compare the obtained magnitude of the dimensionless electron-phonon coupling  with previous theoretical and experimental estimates for the referential lithium-decorated graphene. The representative first-principles calculations for this material yield a total coupling of $\lambda=0.61$, with an appreciable contribution of about $0.1$ arising specifically from the high-energy carbon-carbon stretching modes \cite{profeta2012}. This result was subsequently supported experimentally, with $\lambda=0.58\pm0.05$ and $\sim0.14\pm0.05$ attributed to the high-energy carbon sector \cite{ludbrook2015}. As such, these values provide a useful reference for the value of $\lambda\sim0.05$ presented here, which describes only the effective contribution of the dominant $A_1$ in-plane bond-stretching mode, rather than the complete high-energy carbon phonon sector. In detail, the former comprises all in-plane carbon vibrations, as well as the corresponding dispersive optical branches throughout the Brillouin zone. Therefore, the obtained $\lambda$ describes only a partial contribution to the broader high-energy coupling reported theoretically and experimentally, in accordance with the single-mode character of the present approximation.

To this end, it is argued here that the Kekul{\' e} order may lead to yet another important but nuanced effect associated with the electron-phonon properties of distorted graphene at the van Hove filling. In particular, as already shown, the discussed defect-induced distortion leads to the separation of a well-defined conduction band with a width below the VHS given by $W$. Interestingly, however, closer inspection reveals that, at the fixed filling, Fermi energy remains essentially {\it pinned} to the VHS ($E_{F} \simeq E_{\rm VHS}$) throughout the considered range of Kekul{\' e} order. Hence, $W$ appears to also represent the energy separation between $E_F$ and the bottom of the conduction band, becoming its characteristic electronic energy scale. Importantly, according to Migdal's theorem, such scale should remain large in metals compared with the corresponding phonon energy ($\Omega_{0}$) \cite{migdal1958}:
\begin{equation}
\label{eq11}
\frac{\lambda\Omega_{0}}{W}\ll1.
\end{equation}
Yet, as already observed, the increasing magnitude of the Kekul{\' e} order compresses the conduction band in energy and therefore causes its progressive narrowing by tens of \%, potentially increasing ratio given by Eq. (\ref{eq11}). In other words, the Kekul{\' e} order appears to drive graphene toward the so-called nonadiabatic regime \cite{grimaldi1999}, where the separation between the characteristic electronic and phononic energy scales becomes insufficient for the conventional Migdal approximation to remain well-controlled. In this regime, higher-order electron-phonon scattering processes, can no longer be neglected, such that the effective electron-phonon interaction acquires an explicit dependence on the momentum and frequency structure of the scattering and may be either enhanced or suppressed with respect to its adiabatic description \cite{grimaldi1999}. While the present analysis alone does not establish the onset of such a regime, the observed reduction of the electronic energy scale clearly points in this direction. This possibility is particularly relevant since nonadiabatic corrections were already found to play a significant role in the representative lithium-decorated graphene, notably modifying its superconducting properties \cite{szczesniak2019}.

\section{Conclusions and perspectives}

In the present study, the coexistence of the Kekul{\' e} ordering, flat band formation, and the resulting enhancement of electron-phonon coupling in distorted graphene is theoretically explained within a unified framework. This is done by recalling the bond-resolved picture of the electron-phonon coupling that naturally arises from the structural modification related to the Kekul{\' e} ordering. As a result, the pivotal quantities of interest, such as the conduction band width, the VHS associated with the conduction band, and finally the dimensionless electron-phonon coupling, can all be related to the band gap proxy representing the magnitude of the Kekul{\' e} order in graphene. Taken together, the present results clarify why the increasing Kekul{\' e} order simultaneously favors band flattening and enhanced electron-phonon coupling in graphene, linking both effects to the underlying lattice instability manifested through the distortion-induced differentiation of carbon bonds and accompanying modification of electronic delocalization.

In detail, it is found that the conduction band around the $M$ point indeed exhibits an increasingly flat character that extends over a larger region of momentum space as the Kekul{\' e} order strengthens. At the same time, the VHS peak and the electron-phonon coupling both exhibit approximately quadratic responses to the distortion, with the effective deformation potential remaining only weakly affected. This points to the conclusion that the density of states not only predominantly shapes the electron-phonon coupling in the distorted graphene but also provides a potentially effective route for engineering it. Importantly, the obtained results are shown to be in qualitative agreement with previous theoretical and experimental studies on distorted graphene \cite{profeta2012, ludbrook2015}. Here, of particular importance is lithium-decorated graphene, which has been shown to host the Kekul{\' e} phase accompanied by an extended flat band and increased electron-phonon coupling \cite{bao2022}. Specifically, the present results provide an important microscopic perspective on this experimentally observed correlation, demonstrating a deep interconnection between the strengthening of the Kekul{\' e} order and the promotion of conduction band flattening, as rooted in the mentioned associated bond differentiation and modification of electronic delocalization. In this sense, they partially address the open question raised in \cite{bao2022} concerning nature of the interplay between the flat band and Kekul{\' e} order.

Finally, the presented results appear to have potentially wider implications for the character of the electron-phonon coupling in distorted graphene. This is to say, the progressive narrowing of the conduction band with the strengthening of the Kekul{\' e} order suggests conditions under which the adiabatic picture of Migdal's theorem for metals may be less justified in the considered case. Consequently, the emergence of nonadiabatic effects related to the higher-order electron-phonon scattering processes may become increasingly important. Interestingly, this finding is in line with the earlier observations of such phenomena again in the case of lithium-decorated graphene \cite{szczesniak2019}. In what follows, the established interplay between Kekul{\' e} ordering, flat-band formation, and electron-phonon coupling provides a route toward exploring emerging correlated quantum phases near the VHS. Of particular interest may be scenarios beyond the conventional adiabatic picture, related to other graphene-like flat-band materials with additional structural degrees of freedom ({\it e.g.} bilayer lithium- or calcium-intercalated graphene \cite{guzman2014, wang2022}) or with alternative symmetry distortions such as the sublattice symmetry breaking scheme ({\it e.g.} heavily nitrogen- or boron-doped graphene \cite{zhou2015}).

\bibliography{manuscript}

\end{document}